\documentclass[10pt,conference,letterpaper]{IEEEtran}
\IEEEoverridecommandlockouts

\usepackage{amsmath,amssymb,amsfonts}
\usepackage{algorithm}
\usepackage{algorithmic}
\usepackage{graphicx}
\usepackage{textcomp}
\usepackage{xcolor}
\usepackage{cite}
\usepackage{array}
\usepackage{booktabs}
\usepackage{tabularx}
\usepackage{makecell}
\usepackage{bm}
\usepackage{gensymb}
\usepackage{stfloats}
\usepackage{ragged2e}
\newcommand{\rev}[1]{#1}

\newcolumntype{Y}{>{\RaggedRight\arraybackslash}X}
\newcolumntype{P}[1]{>{\RaggedRight\arraybackslash}p{#1}}

\allowdisplaybreaks[2]

\begin{document}

\title{Learning-Based Beam Adaptation for Active/Passive-Aware Monopulse ISAC with Leakage Suppression}

\author{Jafar~Norolahi,~Alireza~Vahid%
\thanks{J. Norolahi and A. Vahid are with the Department of Electrical and Microelectronic Engineering at Rochester Institute of Technology, Rochester, NY 14623, USA. Emails: \{jn3211@rit.edu, arveme@rit.edu\}.}%
}

\maketitle
\thispagestyle{empty}
\pagestyle{empty}

\begin{abstract}
This paper proposes a learning-based beam adaptation framework for
active/passive-aware monopulse integrated sensing and communication
(ISAC) with spatial leakage suppression. A single gNB uses a shared
array and orthogonal frequency division multiplexing dual-functional
radar-communication signaling to jointly support downlink transmission
and monopulse sensing. Radar detections and uplink spatial evidence are
fused to distinguish cooperative active directions from passive
directions, while a protected angular zone limits unintended
illumination. Soft Actor-Critic deep reinforcement learning adapts the
transmit beam using communication, sensing, leakage, and previous-action
feedback. Numerical results show accurate active/passive spatial
classification and robust leakage control as the number of passive
targets increases from one to twelve. The proposed SAC controller
maintains aggregate passive leakage below the prescribed budget across
the full sweep and, from two passive targets onward, achieves the
highest leakage-compliant sum rate among the considered baselines.
\end{abstract}

\section{Introduction}

Sixth generation (6G) wireless systems are expected to support not only reliable high-rate communications, but also environment awareness through integrated sensing and communication (ISAC). By sharing hardware resources, time frequency resources, and signal structures, ISAC can improve spectral efficiency and reduce implementation overhead compared with separately designed sensing and communication platforms \cite{11231065}. This capability is especially attractive in dynamic wireless environments where the same infrastructure must deliver data services, identify spatial activity, and localize objects or users with limited additional complexity. In such settings, a practical challenge is to continuously adapt the transmit beam in response to changing scene conditions while preserving both communication quality and sensing reliability \cite{10158711,9724187, Norolahi2022SpatialHarmonics}.

Recent studies have established important foundations for this problem from several directions. Cooperative and cell-free ISAC architectures have shown that joint beamforming and distributed processing can improve communication and localization performance in multiuser and multitarget settings \cite{11215603,11288092}. At the same time, secure and privacy aware beamforming formulations have highlighted the importance of controlling unintended radiation toward vulnerable or undesired directions \cite{10605793,11202391}. Learning based methods are also gaining attention because they offer adaptive control policies for large and time varying wireless systems where model based optimization may be difficult to execute online \cite{11345189,11099508,Norolahi2023MLAdaptation}. These developments indicate that sensing aware beam control is becoming a central requirement for practical ISAC deployment.

Despite this progress, a clear gap remains between spatial sensing and closed loop beam adaptation in single-node ISAC systems. Many existing works emphasize either network-level cooperation, waveform design, or secure transmission, but fewer studies integrate radar-based spatial awareness, identification of active and passive directions, and real-time beam adaptation within one unified control loop \cite{11215603,11288092,10605793}. This gap becomes more important when the system must also protect a designated angular sector from unnecessary illumination. It is also important to exploit monopulse sensing, since it enables accurate angle tracking and supports three-dimensional localization with a single node and a single pulse, avoiding the need for multi-node sensing architectures. In such a setting, the transmitter should not only support legitimate users and maintain sensing performance, but should also learn to suppress leakage toward passive directions and a protected zone under changing propagation and user geometry.

Motivated by this need, this paper develops a single-gNB monopulse ISAC framework with learning-based beam adaptation for active/passive-aware spatial control. The proposed design uses sensing feedback to distinguish directions associated with cooperative uplink activity from directions that appear only through radar observations, and exploits this information to update the downlink beam policy over time. By combining monopulse sensing with adaptive beam control, the framework supports accurate localization and tracking while limiting unintended spatial exposure. A protected-zone leakage term is incorporated into the learning objective so that the controller simultaneously promotes useful service and suppresses undesired illumination. In this way, the proposed approach links monopulse sensing, communication-aware beam adaptation, and spatial leakage suppression in one closed-loop architecture.

The main contributions of this paper are as follows:
\begin{itemize}
\item We develop a single-gNB monopulse ISAC framework that jointly supports communication, spatial sensing, accurate localization and tracking, and adaptive beam control in a unified closed-loop architecture.
\item We introduce an active/passive-aware beam adaptation strategy that uses fused uplink and radar evidence to guide learning-based transmit-beam updates.
\item We incorporate protected-zone leakage suppression into the control objective and show that the resulting policy can reduce unintended illumination while maintaining communication and localization performance.
\end{itemize}

\section{System Model and Problem Formulation}
\label{sec:system_model_problem}

This section defines the single-gNB monopulse ISAC setting, the scene geometry of active and passive directions, the protected zone model, and the beam adaptation objective used in the proposed framework.



\begin{figure}[htbp]
  \centering
  \includegraphics[width=.8\columnwidth]{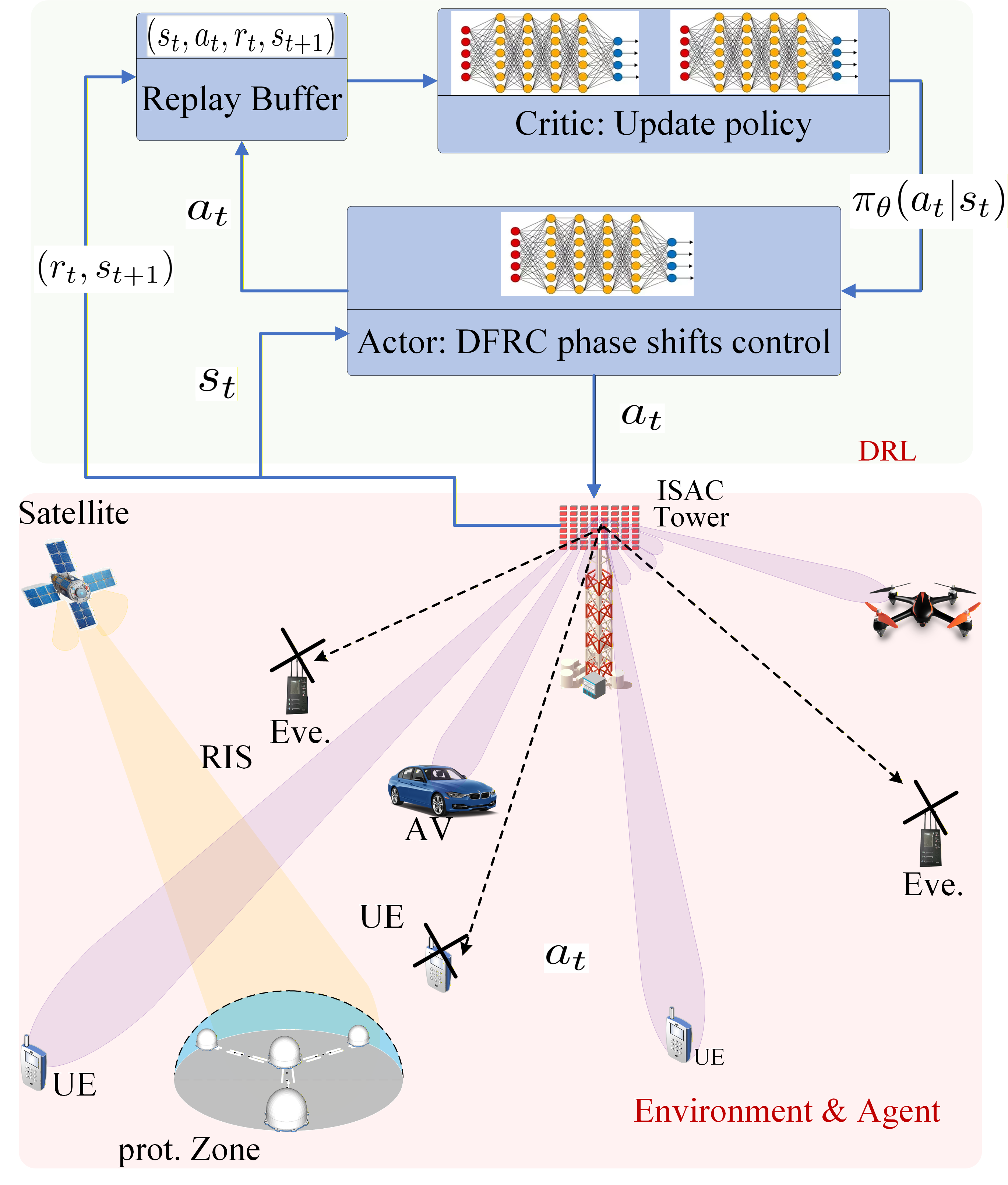}
\caption{Overall architecture of the proposed SAC-DRL-based ISAC beam-control framework.}
\label{fig:ISAC_Idea_conf}
\end{figure}

\subsection{Scenario and Geometry}
\label{subsec:scenario_geometry}

As shown in Fig.~\ref{fig:ISAC_Idea_conf}, we consider a single gNB that jointly performs communication and sensing using a shared antenna platform for downlink transmission and monopulse observation, consistent with ISAC operation in which communication and radar functions reuse common hardware and signaling resources \cite{11231065,10158711,9724187}. The gNB serves legitimate user equipments while monitoring the surrounding scene to identify directions that should be supported or suppressed.

The gNB employs a uniform rectangular array with $N=N_xN_y$ antenna elements. All objects are represented in a three-dimensional Cartesian coordinate system. The gNB position is denoted by $\mathbf{r}_{\mathrm{gNB}}\in\mathbb{R}^{3\times 1}$, and the set of legitimate users is $\mathcal{U}=\{1,\ldots,U\}$, where user $u$ is located at $\mathbf{c}_u\in\mathbb{R}^{3\times 1}$. In addition to these cooperative users, the scene may contain noncooperative reflectors or listeners that generate radar returns without valid uplink activity. These are treated as passive directions and indexed by $\mathcal{P}=\{1,\ldots,P\}$.

For any point $\mathbf{c}\in\mathbb{R}^{3\times 1}$, the direction relative to the gNB is
\begin{align}
\mathbf{d}(\mathbf{c}) = \mathbf{c} - \mathbf{r}_{\mathrm{gNB}},
\end{align}
with azimuth and elevation angles denoted by $\theta(\mathbf{c})$ and $\varphi(\mathbf{c})$. These angles provide the common spatial representation for beamforming, monopulse sensing, and leakage evaluation. An \emph{active direction} corresponds to a legitimate uplink-supported user, whereas a \emph{passive direction} is observed by radar without legal uplink support.
To limit unintended illumination, the protected zone is defined as
\begin{align}
\Omega_Z = \left\{(\theta,\varphi)\,:\,\theta \in [\theta_{\min},\theta_{\max}],\;\varphi \in [\varphi_{\min},\varphi_{\max}] \right\},
\end{align}
where radiation must remain low \cite{11202391,10605793}. The beam adaptation policy therefore aims to support active users while suppressing leakage toward passive directions and the protected zone.




\subsection{Signal and Sensing Model}
\label{subsec:signal_sensing_model}

We consider an orthogonal frequency division multiplexing (OFDM) based dual functional radar communication (DFRC) transmission in which the gNB uses a shared aperture and radio frequency chain for downlink communication and monopulse sensing \cite{11231065,10158711,9724187}. On subcarrier $n\in\{1,\ldots,N_s\}$ and OFDM symbol $\ell\in\{1,\ldots,L\}$, the transmitted signal is
\begin{align}
\mathbf{X}_n[\ell]
=
\mathbf{W}_n \mathbf{s}_n[\ell]
+
\mathbf{r}_n[\ell]
\in \mathbb{C}^{N\times 1},
\label{eq:ofdm_tx_conf}
\end{align}
where $\mathbf{W}_n\in\mathbb{C}^{N\times U}$ is the downlink precoder, $\mathbf{s}_n[\ell]\in\mathbb{C}^{U\times 1}$ is the user symbol vector with $\mathbb{E}\{\mathbf{s}_n[\ell]\mathbf{s}_n[\ell]^H\}=\mathbf{I}_U$, and $\mathbf{r}_n[\ell]\in\mathbb{C}^{N\times 1}$ is the sensing component. The average gNB transmit power satisfies
\begin{align}
\frac{1}{L N_s}\sum_{\ell=1}^{L}\sum_{n=1}^{N_s}
\mathbb{E}\{
\|\mathbf{X}_n[\ell]\|_2^2
\}
\leq
P_{\mathrm{gNB}}.
\label{eq:power_dl_ofdm_conf}
\end{align}

\rev{For sensing, after removing the known symbol on reserved radar resource elements, the received response is}
\begingroup
\begin{align}
\mathbf{y}^{\mathrm{rad}}_n[p]
=
\sum_{q=1}^{Q}
\alpha_q
e^{-j2\pi n\Delta f\tau_q}
e^{j2\pi \nu_q p T_r}
\mathbf{a}_g(\theta_q,\phi_q)
+
\mathbf{n}^{\mathrm{rad}}_n[p],
\label{eq:rad_snapshot_conf}
\end{align}
where $\tau_q=2R_q/c$. The matched-filter range-Doppler map is
\begin{align}
\mathcal M(R,\nu)=\left|\sum_{n\in\mathcal N_R}\sum_p \mathbf w_\Sigma^H\mathbf y_n^{\rm rad}[p]
e^{j4\pi n\Delta fR/c}e^{-j2\pi\nu pT_r}\right|^2,
\label{eq:rd_conf}
\end{align}
whose peaks yield $(\widehat R_l,\widehat\nu_l)$. The monopulse processor forms the sum, azimuth difference, and elevation difference outputs as
\begin{align}
z_{\Sigma}[p]
&=
\mathbf{w}_{\Sigma}^{H}\mathbf{y}^{\mathrm{rad}}_n[p],
\label{eq:mono_sum_conf}
\\
z_{\Delta,\theta}[p]
&=
\mathbf{w}_{\Delta,\theta}^{H}\mathbf{y}^{\mathrm{rad}}_n[p],
\label{eq:mono_diff_az_conf}
\\
z_{\Delta,\phi}[p]
&=
\mathbf{w}_{\Delta,\phi}^{H}\mathbf{y}^{\mathrm{rad}}_n[p],
\label{eq:mono_diff_el_conf}
\end{align}
where the ratio corrections are $\widehat\theta_l=\theta_{0,l}+\Re\{z_{\Delta,\theta}/(z_\Sigma+\epsilon)\}/\kappa_{\theta,l}$ and $\widehat\phi_l=\phi_{0,l}+\Re\{z_{\Delta,\phi}/(z_\Sigma+\epsilon)\}/\kappa_{\phi,l}$, with calibrated local slopes $\kappa_{\theta,l},\kappa_{\phi,l}$. Finally, $\widehat{\mathbf p}_l=\mathbf p_g+\widehat R_l[\cos\widehat\phi_l\cos\widehat\theta_l,\cos\widehat\phi_l\sin\widehat\theta_l,\sin\widehat\phi_l]^T$. The resulting radar detection set is
\begin{align}
\mathcal{D}
=
\{(\widehat R_l,\widehat\nu_l,\widehat{\theta}_l,\widehat{\phi}_l)\}_{l\in\mathcal{L}},
\label{eq:det_set_conf}
\end{align}
which completes the delay, Doppler, monopulse-angle, and 3-D localization chain.
In parallel, the gNB forms an uplink spatial spectrum from the received uplink snapshots. The post fast Fourier transform uplink snapshot on subcarrier $n$ at symbol index $p$ is
\begin{align}
\mathbf{y}^{\mathrm{UL}}_n[p]
=
\sum_{u=1}^{U}
\mathbf{h}_{u,g}[n]\,
s^{\mathrm{UL}}_{u,n}[p]
+
\mathbf{n}_n[p],
\label{eq:ul_snap_conf}
\end{align}
where $\mathbf{h}_{u,g}[n]\in\mathbb{C}^{N\times 1}$ is the uplink channel of user $u$, $s^{\mathrm{UL}}_{u,n}[p]$ is the uplink symbol, and $\mathbf{n}_n[p]$ is receiver noise. Using
\begin{align}
\mathbf{b}(\theta,\phi)
=
\frac{\mathbf{a}_g(\theta,\phi)}{\sqrt{N}},
\label{eq:ul_beamformer_conf}
\end{align}
the wideband uplink spatial spectrum is
\begin{align}
\mathcal{S}(\theta,\phi)
=
\frac{1}{|\mathcal{N}_{\mathrm{UL}}|}
\sum_{n\in\mathcal{N}_{\mathrm{UL}}}
\sum_{p\in\mathcal{T}_{\mathrm{UL}}}
\left|
\mathbf{b}(\theta,\phi)^H
\mathbf{y}^{\mathrm{UL}}_n[p]
\right|^2,
\label{eq:ul_spectrum_conf}
\end{align}
where $\mathcal{N}_{\mathrm{UL}}$ is the selected uplink subcarrier set and $\mathcal{T}_{\mathrm{UL}}$ is the temporal observation window.

The radar detections and uplink spectrum are fused to identify active and passive directions. The active direction set is
\begin{align}
\mathcal{A}
=
\{&
(\widehat{\theta}_l,\widehat{\phi}_l)\in\mathcal{D}
:
\max_{(\theta,\phi)\in\mathcal{B}(\widehat{\theta}_l,\widehat{\phi}_l)}
\mathcal{S}(\theta,\phi)
\geq
\tau_{\mathrm{UL}}
\},
\label{eq:active_set_conf}
\end{align}
where $\mathcal{B}(\widehat{\theta}_l,\widehat{\phi}_l)$ is a local angular gate and $\tau_{\mathrm{UL}}$ is the uplink support threshold. The passive direction set is
\begin{align}
\mathcal{E}
=
\mathcal{D}\setminus\mathcal{A},
\label{eq:eve_set_conf}
\end{align}
which contains radar detected directions without sufficient uplink support. The sets $\mathcal{A}$ and $\mathcal{E}$ are then used by the leakage aware beam adaptation objective together with the protected zone.



\subsection{Design Objective}
\label{subsec:design_objective}

Using the active-direction set $\mathcal{A}$, the passive-direction set $\mathcal{E}$, and the protected zone $\Omega_Z$ defined in the previous subsection, we now specify the beam adaptation objective. The design seeks to maintain reliable service for the legitimate users while suppressing unintended downlink radiation toward passive directions and the protected zone, which is consistent with privacy-aware and secure ISAC formulations \cite{10605793,11202391}.

For user $u\in\mathcal{U}$, the received signal on subcarrier $n$ and OFDM symbol $\ell$ is
\begin{align}
y_{n,u}[\ell]
&=
\mathbf{h}_{g,u}[n]^{H}\mathbf{X}_n[\ell]
+
z_{n,u}[\ell]
\nonumber\\
&=
\mathbf{h}_{g,u}[n]^{H}\mathbf{W}_n\mathbf{s}_n[\ell]
+
\mathbf{h}_{g,u}[n]^{H}\mathbf{r}_n[\ell]
+
z_{n,u}[\ell],
\label{eq:rx_user_conf}
\end{align}
where $\mathbf{h}_{g,u}[n]\in\mathbb{C}^{N\times 1}$ denotes the downlink channel from the gNB to user $u$ on subcarrier $n$, and $z_{n,u}[\ell]\sim\mathcal{CN}(0,\sigma_u^2)$. Accordingly, the per-subcarrier SINR of user $u$ is
\begin{align}
\gamma_{n,u}
=
\frac{
\left|\mathbf{h}_{g,u}[n]^{H}\mathbf{w}_{n,u}\right|^2
}{
\sum_{v\neq u}\left|\mathbf{h}_{g,u}[n]^{H}\mathbf{w}_{n,v}\right|^2
+
\left|\mathbf{h}_{g,u}[n]^{H}\mathbf{r}_n[\ell]\right|^2
+
\sigma_u^2
},
\label{eq:sinr_user_conf}
\end{align}
where $\mathbf{W}_n=[\mathbf{w}_{n,1},\ldots,\mathbf{w}_{n,U}]$. A compact communication-performance metric is then defined through the average achievable rate
\begin{align}
\bar{R}_u
=
\frac{1}{N_s}\sum_{n=1}^{N_s}\log_2\!\left(1+\gamma_{n,u}\right),
\qquad u\in\mathcal{U},
\label{eq:avg_rate_user_conf}
\end{align}
and the system communication utility is measured by the sum rate
\begin{align}
R_{\mathrm{sum}}
=
\sum_{u\in\mathcal{U}}\bar{R}_u.
\label{eq:sum_rate_conf}
\end{align}
\rev{Next, we quantify the unintended downlink leakage toward the passive-direction set $\mathcal{E}$. For each detected passive direction $(\widehat{\theta}_l,\widehat{\phi}_l)\in\mathcal{E}$, let $\mathbf{a}_g(\widehat{\theta}_l,\widehat{\phi}_l)\in\mathbb{C}^{N\times 1}$ denote the corresponding gNB steering vector. To match the implementation, leakage is evaluated by the worst sensed direction rather than by averaging over directions:}
\begin{align}
\mathcal{L}_{E}
=
\max_{(\widehat{\theta}_l,\widehat{\phi}_l)\in\mathcal{E}}
\frac{1}{N_s}
\sum_{n=1}^{N_s}
\left\|
\mathbf{a}_g(\widehat{\theta}_l,\widehat{\phi}_l)^{H}\mathbf{W}_n
\right\|_2^2.
\label{eq:leak_passive_conf}
\end{align}
\rev{This conservative metric measures the strongest communication-beam exposure over the radar-detected directions that are not supported by valid uplink activity.}

\rev{To protect the designated angular sector, we also define the worst sampled leakage toward the protected zone $\Omega_Z$. Let $\{(\theta_j^{Z},\phi_j^{Z})\}_{j=1}^{J}\subset\Omega_Z$ denote a set of angular samples drawn from the protected zone. The protected-zone leakage is}
\begin{align}
\mathcal{L}_{Z}
=
\max_{1\leq j\leq J}
\frac{1}{N_s}
\sum_{n=1}^{N_s}
\left\|
\mathbf{a}_g(\theta_j^{Z},\phi_j^{Z})^{H}\mathbf{W}_n
\right\|_2^2.
\label{eq:leak_zone_conf}
\end{align}
Therefore, $\mathcal{L}_{E}$ and $\mathcal{L}_{Z}$ jointly capture the two undesired spatial-radiation mechanisms considered in this work, namely passive-direction exposure and protected-zone exposure.

To keep the formulation compact, sensing support is enforced implicitly through the retained DFRC sensing signal and its power allocation, rather than through an additional explicit localization term in the objective. Accordingly, the beam adaptation design is posed as
\begin{align}
\max_{\{\mathbf{W}_n\},\,\{\mathbf{r}_n[\ell]\}}
\quad
R_{\mathrm{sum}}
\label{eq:beam_opt_conf}
\end{align}
subject to
\begin{align}
&\frac{1}{L N_s}\sum_{\ell=1}^{L}\sum_{n=1}^{N_s}
\mathbb{E}\!\left[
\left\|
\mathbf{W}_n\mathbf{s}_n[\ell]+\mathbf{r}_n[\ell]
\right\|_2^2
\right]
\leq
P_{\mathrm{gNB}},
\tag{\ref{eq:beam_opt_conf}a}\label{eq:cons_total_power_conf}
\\
&\frac{1}{L N_s}\sum_{\ell=1}^{L}\sum_{n=1}^{N_s}
\mathbb{E}\!\left[
\left\|
\mathbf{r}_n[\ell]
\right\|_2^2
\right]
\leq
P_{\mathrm{rad}},
\tag{\ref{eq:beam_opt_conf}b}\label{eq:cons_radar_power_conf}
\\
&\mathcal{L}_{E}\leq L_{E}^{\max},
\qquad
\mathcal{L}_{Z}\leq L_{Z}^{\max},
\tag{\ref{eq:beam_opt_conf}c}\label{eq:cons_leak_conf}
\\
&\bar{R}_u\geq R_u^{\min},
\qquad \forall u\in\mathcal{U},
\tag{\ref{eq:beam_opt_conf}d}\label{eq:cons_rate_conf}
\end{align}
where $P_{\mathrm{gNB}}$ is the total gNB transmit-power budget, $P_{\mathrm{rad}}$ is the sensing-power budget, and $L_{E}^{\max}$ and $L_{Z}^{\max}$ are the admissible leakage thresholds for the passive directions and the protected zone, respectively. The optional rate constraint in \eqref{eq:cons_rate_conf} preserves a minimum quality of service for the legitimate users.
Problem \eqref{eq:beam_opt_conf} is nonconvex because the user SINRs are coupled through multiuser interference and because the communication and sensing components share the same transmit resources.



\section{Proposed Learning-Based Beam Adaptation via SAC-DRL}
\label{sec:proposed_sac}

Problem \eqref{eq:beam_opt_conf} is challenging to solve online because the environment changes over time and the communication and leakage objectives are tightly coupled. Therefore, we use SAC-DRL to learn an adaptive beam-control policy from communication and sensing feedback.

\subsection{MDP Formulation}
\label{subsec:mdp_formulation}

Beam adaptation is modeled as a Markov decision process (MDP) with one decision per frame $t$. At frame $t$, the controller observes state $\mathbf{s}_t$, selects action $\mathbf{a}_t$, applies it at the gNB, and receives reward $r_t$ and next state $\mathbf{s}_{t+1}$:
\begin{align}
\mathbf{s}_t
\;\xrightarrow{\;\mathbf{a}_t\;}\;
r_t,\;\mathbf{s}_{t+1}.
\label{eq:mdp_transition_conf}
\end{align}
The policy $\pi(\mathbf{a}_t|\mathbf{s}_t)$ is learned to maximize
\begin{align}
J(\pi)
=
\mathbb{E}_{\pi}
\left[
\sum_{t=0}^{\infty}
\gamma^{t} r_t
\right],
\qquad 0<\gamma<1,
\label{eq:mdp_return_conf}
\end{align}
where $\gamma$ is the discount factor. This formulation casts beam adaptation as a sequential decision problem that balances communication performance and leakage suppression.



\subsection{State Design}
\label{subsec:state_design}

\rev{The implemented state uses only quantities available to the controller: up to four normalized active azimuth/elevation pairs, up to four sensed passive pairs, the protected-zone center and current SNR, four normalized user-rate feedback entries, the normalized passive/zone violation magnitudes, and the previous action:}
\begin{align}
\mathbf{s}_t
=
\Big[
\mathbf{f}_{\mathcal{A},t}^{T},
\mathbf{f}_{\mathcal{E},t}^{T},
\bar{\mathbf{z}}_{t}^{T},
\xi_t,
\mathbf{r}_{t}^{T},
v_{E,t},v_{Z,t},
\mathbf{a}_{t-1}^{T}
\Big]^{T}.
\label{eq:state_vector_conf}
\end{align}
\rev{Here $\bar{\mathbf{z}}_t$ is the protected-zone angular center, $\xi_t$ is the SNR feature, and $v_{E,t},v_{Z,t}$ are the leakage-budget violations defined below. No ground-truth passive direction is included in the state; true directions are used only to score the held-out evaluation.}



\subsection{Action Design}
\label{subsec:action_design}

The learning action is devoted entirely to gNB beam adaptation. At frame $t$, the SAC-DRL policy outputs a continuous action vector
\begin{align}
\mathbf{a}_t
\in
\mathbb{R}^{d_a},
\label{eq:action_vector_conf}
\end{align}
which is mapped to a feasible transmit configuration through
\begin{align}
\{\mathbf{W}_n^{(t)}\}_{n=1}^{N_s}
=
\Psi(\mathbf{a}_t),
\label{eq:action_projection_conf}
\end{align}
where $\Psi(\cdot)$ enforces the transmit-power constraints and beamformer structure. 
This lower-dimensional parameterization keeps the action space tractable while allowing adaptive control of the effective gNB beam and, if needed, the communication-versus-sensing power allocation.



\subsection{Reward Design}
\label{subsec:reward_design}

\rev{The reward promotes communication performance while penalizing only leakage-budget violations, rather than rewarding unnecessary over-nulling. Let $\mathcal{L}_{E,t}^{\mathrm{dB}}$ and $\mathcal{L}_{Z,t}^{\mathrm{dB}}$ denote the dB versions of \eqref{eq:leak_passive_conf} and \eqref{eq:leak_zone_conf}, and define}
\begin{align}
v_{E,t}&=\max\{0,\mathcal{L}_{E,t}^{\mathrm{dB}}-L_{E,\max}^{\mathrm{dB}}\},\nonumber\\
v_{Z,t}&=\max\{0,\mathcal{L}_{Z,t}^{\mathrm{dB}}-L_{Z,\max}^{\mathrm{dB}}\}.
\end{align}
\rev{The implemented reward is}
\begin{align}
r_t
=&\; R_{\mathrm{sum},t}
-\lambda_E v_{E,t}-\rho_E v_{E,t}^2
-\lambda_Z v_{Z,t}-\rho_Z v_{Z,t}^2 \nonumber\\
&-\lambda_{\mathrm{loc}} e_{3D,t}
-\eta_a \left\|\mathbf{a}_t-\mathbf{a}_{t-1}\right\|_2,
\label{eq:reward_conf}
\end{align}
\rev{where $e_{3D,t}$ is the mean 3-D sensing error. The implemented weights, for example, in our study are $(\lambda_E,\rho_E,\lambda_Z,\rho_Z,\lambda_{\mathrm{loc}},\eta_a)=(0.40,0.30,0.35,0.25,0.002,0.01)$. This form penalizes violations without rewarding unnecessary over-nulling, while failure-rate and violation statistics remain separately visible.}



\subsection{SAC-DRL Update Mechanism}
\label{subsec:sac_update}

To learn the adaptive beam-control policy in the continuous action space, we use SAC-DRL with a stochastic actor, twin critics, target critics, and a replay buffer. At each frame, the transition
\begin{align}
(\mathbf{s}_t,\mathbf{a}_t,r_t,\mathbf{s}_{t+1})
\label{eq:transition_tuple_conf}
\end{align}
is stored in the replay buffer $\mathcal{D}$ and sampled for mini-batch training. For each sampled transition, the soft Bellman target is
\begin{align}
y_t
&=
r_t
+
\gamma
\mathbb{E}_{\mathbf{a}_{t+1}\sim\pi_{\phi}}
\Big[
\bar{Q}(\mathbf{s}_{t+1},\mathbf{a}_{t+1})
\nonumber\\
&\qquad
-
\alpha_{\mathrm{ent}}
\log \pi_{\phi}(\mathbf{a}_{t+1}|\mathbf{s}_{t+1})
\Big],
\label{eq:sac_target_conf}
\end{align}
where
\begin{align}
\bar{Q}(\mathbf{s},\mathbf{a})
=
\min_{i\in\{1,2\}}
Q_{\bar{\psi}_i}(\mathbf{s},\mathbf{a}).
\label{eq:sac_qbar_conf}
\end{align}
\rev{The two critics are updated by minimizing the squared Bellman error with respect to $y_t$, and the actor is updated to maximize the entropy-regularized critic value. The implementation uses a fixed entropy weight $\alpha_{\mathrm{ent}}=0.05$ rather than automatic temperature tuning. Training uses stochastic actions; evaluation freezes the actor and applies its deterministic mean action. The actor and twin critics use two 64-unit ReLU layers, a $10^5$-transition replay buffer, batch size 64, discount 0.98, one update per frame, Polyak factor 0.01, five independent seeds, and 140 episodes of 20 frames per seed (2800 transitions per policy), with recorded NumPy seeds.}


\subsection{Closed-Loop Beam Adaptation Procedure}
\label{subsec:closed_loop_procedure}

Algorithm~\ref{alg:closed_loop_beam_adaptation} summarizes the
frame-wise closed loop. At each frame, communication and sensing outputs
form the state, the SAC policy generates a beam-control action, and
$\Psi(\cdot)$ maps it to a feasible transmit configuration. The resulting
transition is stored in the replay buffer and used to update the actor,
twin critics, and target networks.

\begin{algorithm}[t]
\caption{Closed-Loop Learning-Based Beam Adaptation}
\scriptsize
\label{alg:closed_loop_beam_adaptation}
\begin{algorithmic}[1]
\STATE Initialize actor $\pi_{\phi}$, twin critics $Q_{\psi_1},Q_{\psi_2}$,
target critics, and replay buffer $\mathcal{D}$
\FOR{each frame $t$}
    \STATE Collect communication/sensing outputs and form $\mathbf{s}_t$
    \STATE Sample $\mathbf{a}_t\!\sim\!\pi_{\phi}(\cdot|\mathbf{s}_t)$
    and obtain feasible beams through $\Psi(\mathbf{a}_t)$
    \STATE Apply the beams and observe $r_t,\mathbf{s}_{t+1}$
    \STATE Store $(\mathbf{s}_t,\mathbf{a}_t,r_t,\mathbf{s}_{t+1})$
    in $\mathcal{D}$ and sample a mini-batch
    \STATE Update the twin critics and actor using the SAC objectives
    \STATE Soft-update the target critics
\ENDFOR
\end{algorithmic}
\end{algorithm}



\section{Numerical Results}

\rev{Unless otherwise stated, simulations use 28 GHz, a $40\times40$
half-wavelength uniform rectangular array (URA), 128 subcarriers, 14
orthogonal frequency division multiplexing (OFDM) symbols, 16
radar-reserved subcarriers, 64 radar pulses, 100-kHz spacing, 14
protected-zone samples over azimuth $40$-$70^\circ$ and elevation
$-5$-$10^\circ$, and a $2^\circ$ association gate. Passive- and
protected-zone leakage budgets are $-45$ and $-50$ dB. For
Fig.~\ref{fig:robust_rate} and Fig.~\ref{fig:robust_leak}, passive
targets vary from 1 to 12 at 5-dB SNR.}

\rev{The reference geometry is used for
Figs.~\ref{fig:gnb_doa}-\ref{fig:rates_cnf}, whereas
Fig.~\ref{fig:robust_rate} evaluates leakage-compliant sum-rate
scalability and Fig.~\ref{fig:robust_leak} evaluates aggregate-leakage
scalability under randomized geometry and passive-target population.
The baselines are fixed maximum-ratio transmission (Fixed MRT), an
unconstrained beamforming reference; heuristic nulling, which places
explicit nulls toward sensed passive and protected directions; and deep
deterministic policy gradient (DDPG), a deterministic
deep-reinforcement-learning baseline trained in the same beam-control
environment as the proposed Soft Actor-Critic (SAC) controller.}

\begin{figure}[htbp]
  \centering
  \includegraphics[width=.8\columnwidth]{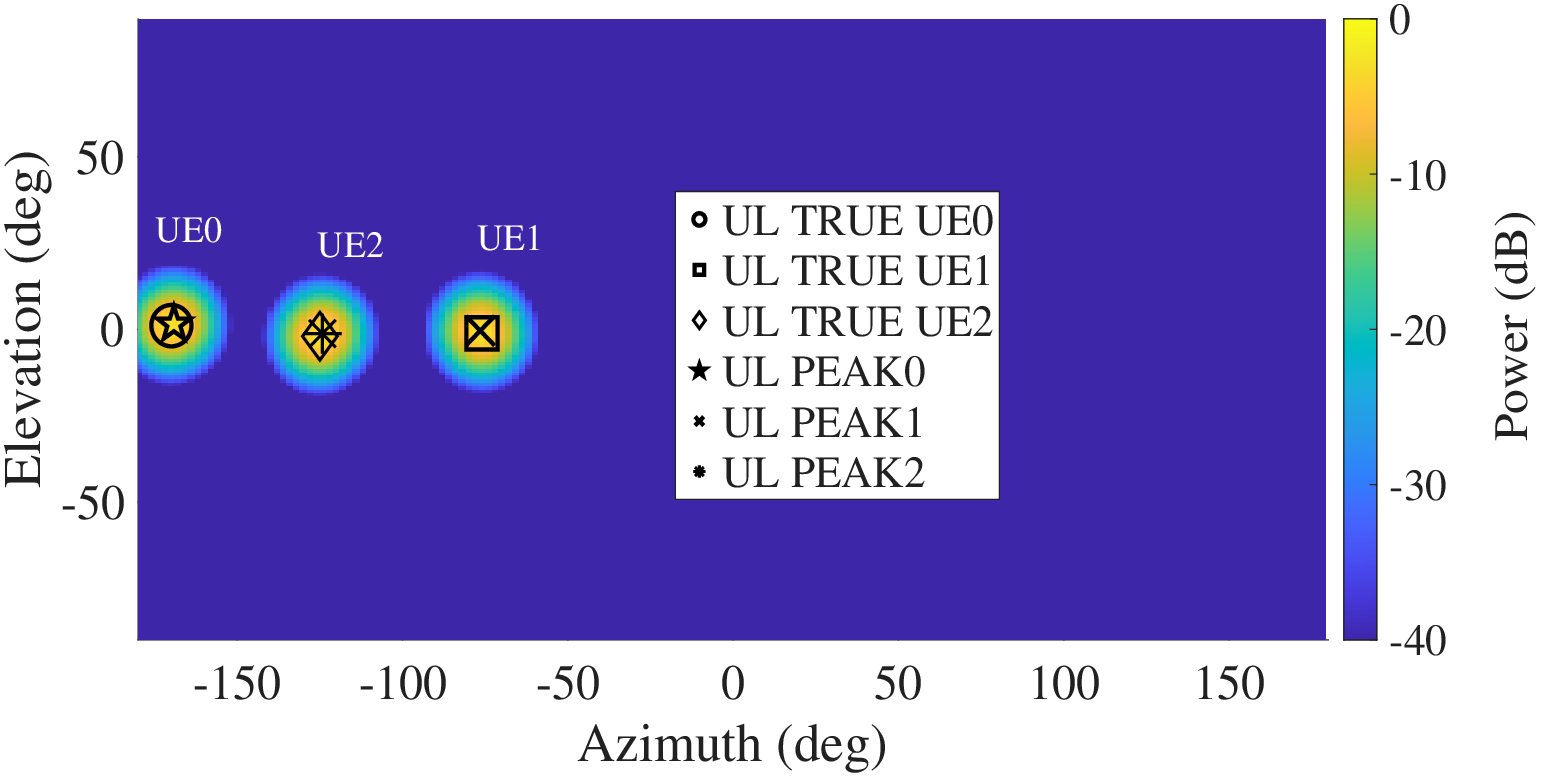}
\caption{Communication-side DOA estimation showing the true uplink
directions and estimated spatial peaks.}
\label{fig:gnb_doa}
\end{figure}

Figure~\ref{fig:gnb_doa} shows direction-of-arrival (DOA) errors of
approximately $0.87^\circ$, $0.92^\circ$, and $1.06^\circ$ for UE0,
UE1, and UE2, respectively. Because UE2 lies near the protected zone,
only UE0 and UE1 are treated as legal downlink users.

\begin{figure}[htbp]
  \centering
  \includegraphics[width=.8\columnwidth]{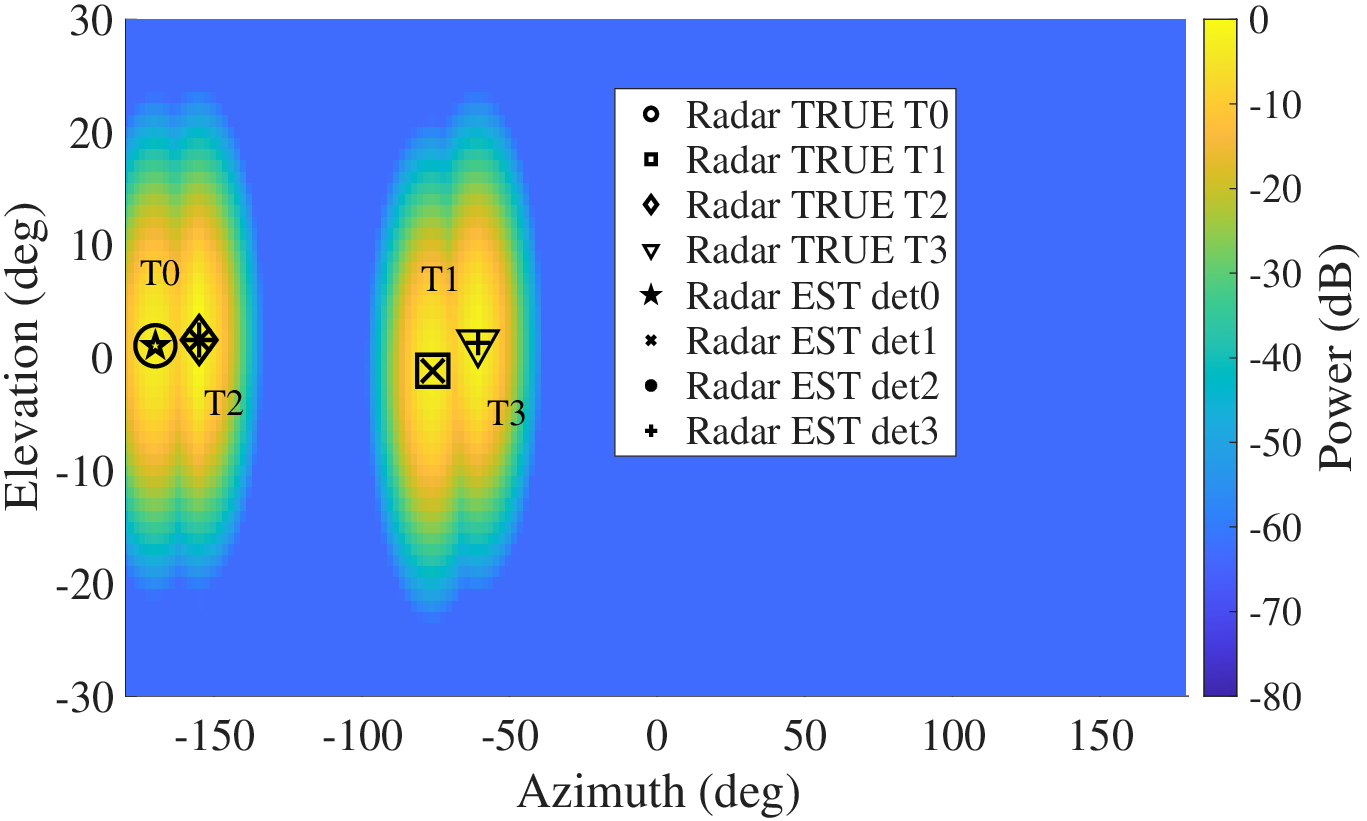}
\caption{Radar-side DOA estimation for the active users and passive
targets.}
\label{fig:sensing_doa}
\end{figure}

Figure~\ref{fig:sensing_doa} gives three-dimensional localization errors
of $0.499$, $0.104$, $0.479$, and $0.067$ m. With the $2^\circ$
association gate, detections 0 and 1 are classified as active, whereas
detections 99 and 100 are classified as passive.

\begin{figure}[htbp]
  \centering
  \includegraphics[width=.8\columnwidth]{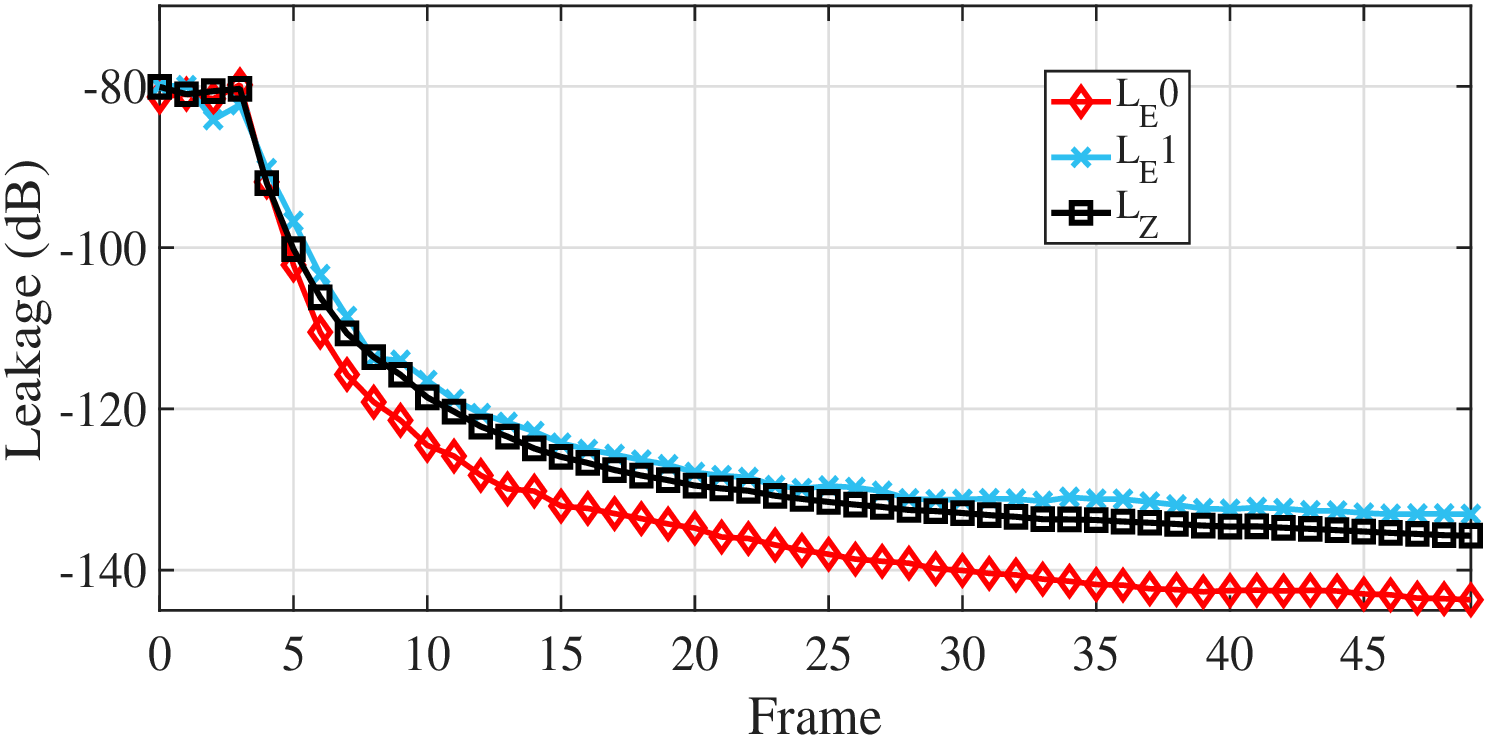}
\caption{Passive-direction and protected-zone leakage over the frames
under the proposed beam adaptation.}
\label{fig:leakages}
\end{figure}

Figure~\ref{fig:leakages} shows progressive suppression: the two
passive-direction leakages decrease from $-81.3/-79.9$ dB to
$-143.7/-133.1$ dB, while protected-zone leakage decreases from
$-80.0$ to $-135.7$ dB.

\begin{figure}[htbp]
  \centering
  \includegraphics[width=.8\columnwidth]{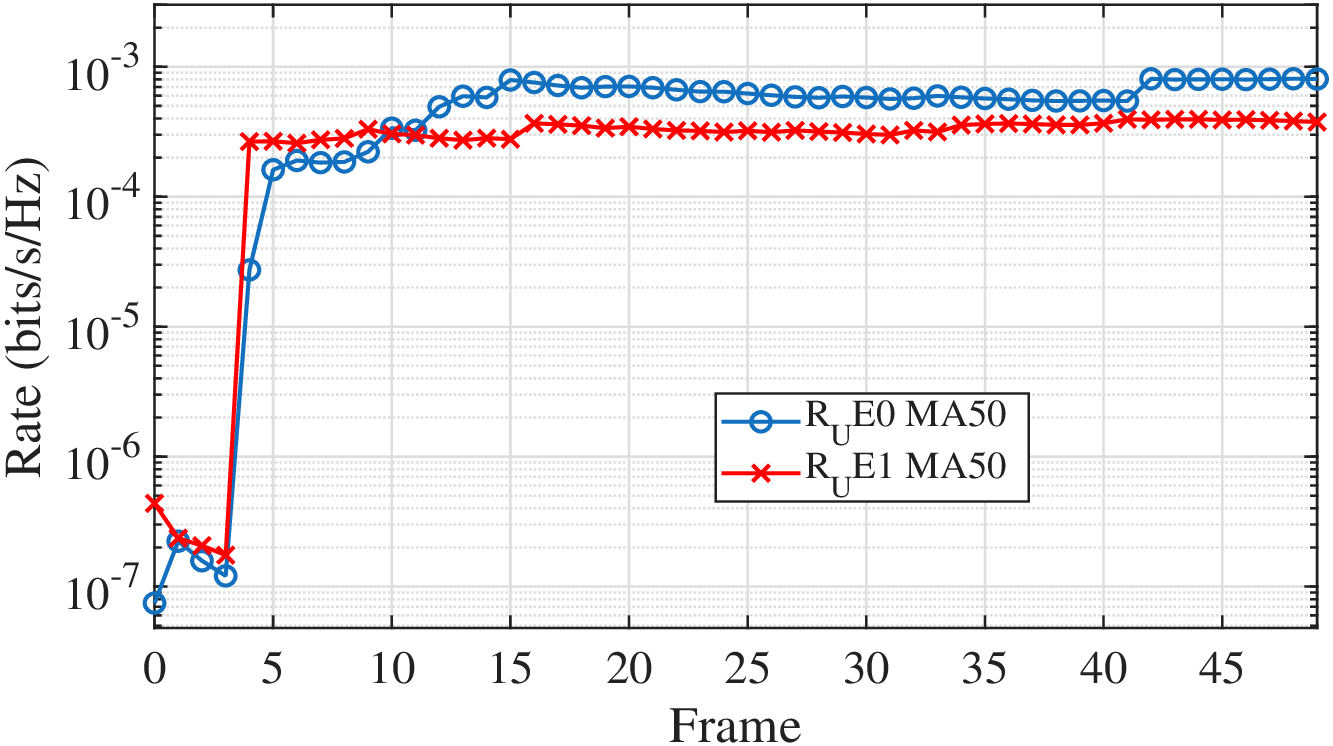}
\caption{Achieved rates of the two legal users over the adaptation
frames.}
\label{fig:rates_cnf}
\end{figure}

Figure~\ref{fig:rates_cnf} shows UE0 reaching
$8.05\times10^{-4}$ bits/s/Hz while UE1 stabilizes near
$3.76\times10^{-4}$ bits/s/Hz, demonstrating legal-user service under
the leakage-aware adaptation.

\begin{figure}[t]
\centering
\includegraphics[width=.8\columnwidth]
{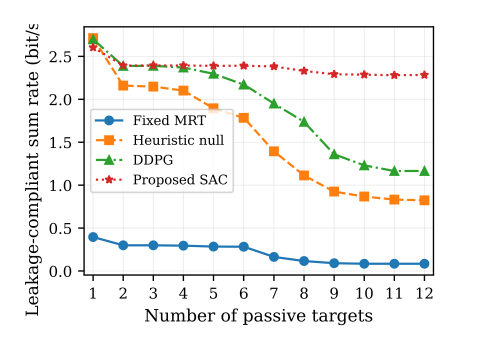}
\caption{Leakage-compliant sum rate versus the number of
passive targets. }
\label{fig:robust_rate}
\end{figure}

\begin{figure}[t]
\centering
\includegraphics[width=.8\columnwidth]
{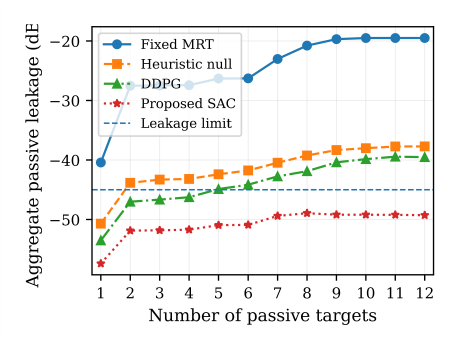}
\caption{Aggregate passive leakage versus the number of
passive targets. The dashed line denotes the $-45$-dB leakage budget.}
\label{fig:robust_leak}
\end{figure}

\rev{At one target, SAC achieves $2.604$ bit/s/Hz versus $2.713$ and
$2.699$ bit/s/Hz for heuristic nulling and DDPG. From two targets
onward, SAC provides the highest leakage-compliant rate in
Fig.~\ref{fig:robust_rate}, remaining near $2.39$ bit/s/Hz for
two-six targets and reaching $2.283$ bit/s/Hz at twelve. At twelve
targets, DDPG and heuristic nulling fall to $1.165$ and $0.823$
bit/s/Hz, respectively, giving SAC approximately $96\%$ and $177\%$
higher feasible rates.}
\rev{SAC maintains aggregate passive leakage below $-45$ dB throughout
the 1-12 target sweep in Fig.~\ref{fig:robust_leak}, from $-57.40$ to
$-49.28$ dB. DDPG crosses the budget at five targets and reaches
$-39.52$ dB at twelve, while heuristic nulling and Fixed MRT reach
$-37.72$ and $-19.50$ dB. Consequently, the baselines require increasing
power backoff, whereas SAC preserves leakage feasibility and usable
communication power.}
\rev{Additional stress tests show that calibration error increases angular
error, missed detections can violate the leakage budget, and mobility
causes a moderate rate reduction, complementing the scalability results
above.}


\section{Conclusion}

This paper developed a single-gNB active/passive-aware monopulse ISAC
framework that combines uplink-supported spatial classification,
radar-based 3-D sensing, protected-zone awareness, and SAC-based
beam adaptation. The proposed controller jointly preserves useful
communication service while limiting unintended illumination toward
passive directions and the protected zone. Numerical results confirm
accurate active/passive discrimination and show that, as the number of
passive targets increases from one to twelve, SAC maintains aggregate
passive leakage below the prescribed budget while preserving a
comparatively stable communication rate. From two passive targets
onward, it achieves the highest leakage-compliant sum rate among Fixed
MRT, heuristic nulling, and DDPG. Additional calibration,
missed-detection, and mobility tests expose the remaining operating
limits of the learned controller.

\bibliographystyle{IEEEtran}
\bibliography{ISAC_Monopulse_RIS_SACDRL}

\end{document}